\documentclass[aps,preprintnumbers,superscriptaddress,showpacs]{revtex4}
\usepackage{epsfig}
\usepackage{psfrag}
\usepackage{amsfonts}
\usepackage{graphicx}
\usepackage{dcolumn}
\usepackage{bm}

\begin{document}

\title{The effect of chemical potential on imaginary potential and entropic force}

\author{Zi-qiang Zhang}
\email{zhangzq@cug.edu.cn} \affiliation{School of mathematics and
physics, China University of Geosciences(Wuhan), Wuhan 430074,
China}

\author{De-fu Hou}
\email{houdf@mail.ccnu.edu.cn} \affiliation{Institute of Particle
Physics and Key Laboratory of Quark and Lepton Physics (MOS),
Central China Normal University, Wuhan 430079,China}

\author{Gang Chen}
\email{chengang1@cug.edu.cn} \affiliation{School of mathematics
and physics, China University of Geosciences(Wuhan), Wuhan 430074,
China}

\begin{abstract}
Imaginary potential and entropic force represent different
mechanisms for melting the heavy quarkonium. In this paper, we
study the chemical potential effect on these two quantities with
respect to a moving quarkonium from the AdS/CTF duality. We
observe that for both mechanisms the chemical potential has the
same effect: the presence of the chemical potential tends to
decrease the dissociation length.

\end{abstract}


\maketitle
\section{Introduction}
The experiments of heavy ion collisions at LHC and RHIC have
produced a new state of matter so-called "strong quark-gluon
plasma(sQGP)" \cite{JA,KA,EV}. One main experimental signatures
for sQGP formation is melting of quarkoniums, such as $J/\psi$ and
excited states, in the medium \cite{KM}. It was argued earlier
\cite{TMA} that the color screening is the main mechanism
responsible for this suppression. But recently some other
mechanisms, such as the imaginary potential \cite{MLA,MAE} and the
entropic force \cite{DEK} have been proposed. AdS/CFT
\cite{Maldacena:1997re,Gubser:1998bc,MadalcenaReview}, the duality
between a string theory in AdS space and a conformal field theory
in the physical space-time, has yielded many important insights
for studying strongly coupled field theory. This method has been
used to study the imaginary potential and the entropic force
recently.

The imaginary potential can be used to estimate the width of heavy
quarkonia in plasma. Also, it can yield to the suppression of the
$Q\bar{Q}$ in heavy ion collisions. By using the AdS/CFT duality,
J. Noronha et al have carried out the imaginary potential for
$\mathcal{N}=4$ SYM theory in their seminal work \cite{JN}. There,
the Im$V_{Q\bar{Q}}$ is related to the effect of thermal
fluctuations due to the interactions between the heavy quarks and
the medium. Soon the investigations of \cite{JN} have been
extended to various cases. For instance, the imaginary potential
of static quarkonium in strongly coupled plasma is discussed in
\cite{JN1,KB}. The Im$V_{Q\bar{Q}}$ of moving quarkonium in
strongly coupled plasma has been studied in \cite{MAL}. The finite
't Hooft coupling corrections on this quantity is analyzed in
\cite{KB1}. Other important results can be found, for example, in
\cite{DGI,JS,JS1}. In addition, there are other approaches which
can also extract an imaginary part of the potential from the
holography \cite{JLA,THA}.

Another important quantity related to the suppression of
$Q\bar{Q}$ is the entropic force. This force has been introduced
long time ago \cite{KH} and proposed to responsible for the
gravity recently \cite{EP}. In a more recent work, D. E. Kharzeev
\cite{DEK} show that the entropic force is responsible for
dissociating the quarkonium. This argument is based upon the
Lattice QCD results that a large amount of entropy related to the
heavy quarknioum placed in the QGP \cite{DKA1,DKA2,PPE}. Applying
the AdS/CFT, K. Hashimoto et al have studied the entropic force
associated with the heavy quark pair firstly \cite{KHA}. It is
shown that the peak of the entropy near the transition point
associates with the nature of deconfinement. After \cite{KHA}, the
entropic force of a moving heavy quarkonium has been studied in
\cite{KBF}, it is found that the velocity leads to increasing the
entropic force thus making the moving quarkonium dissociates
easier. We have investigated the entropic force of a rotating
heavy quarkonium in \cite{ZQ} and observed that the rotating
quarkonium dissociates harder than the static case.

As we know, the $Q\bar{Q}$ pair is not produced at rest in sQGP.
Therefore, the effect of the medium in motion of $Q\bar{Q}$ should
be taken into account. In this paper, we will investigate the
chemical potential effect on the imaginary potential as well as
the entropic force with respect to a moving quarkonium in plasma
from AdS/CFT. We would like to see how the chemical potential
affects the two quantities or the quarkonium dissociation.
Moreover, imaginary potential and entropic force represent
different mechanisms for the melting of the heavy quarkonium, so
it would be interesting to compare them. Evaluations of the
chemical potential effect on these two quantities could be
considered as a simple text of this "comparison". These are the
main motivations of the present work.

We organize the paper as follows. In the next section, we briefly
review the AdS Schwarzchild background with chemical potential and
boost the frame in one direction. In section 3, we study the
imaginary potential of a moving quarkonium in this background and
discuss the effect of the chemical potential on it. The chemical
potential effect on the entropic force with respect to a moving
quarkonium in the same background will be investigated in section
4. The last section is devoted to conclusion and discussion.

\section{Setup}
In the holographic dictionary, $\mathcal N=4$ SYM theory with a
chemical potential can be obtained by making the black hole in the
holographic dimension be charged. The corresponding spacetime is
the $AdS_5$-Reissner-Nordstrom geometry. The metric is given by
\begin{eqnarray}
ds^2
=-\frac{r^2}{R^2}f(r)dt^2+\frac{r^2}{R^2}d\vec{x}^2+\frac{R^2}{r^2}f(r)^{-1}dr^2,\label{metric}
\end{eqnarray}
with
\begin{equation}
f(r)=1-(1+Q^2)(\frac{r_h}{r})^4+Q^2(\frac{r_h}{r})^6,
\end{equation}
where $R$ is the AdS space radius, r denotes the radial coordinate
with $r=r_h$ the horizon. The string tension is
$\frac{1}{2\pi\alpha^\prime}$ where $\alpha^\prime$ is related to
the 't Hooft coupling constant by
$\frac{R^2}{\alpha^\prime}=\sqrt{\lambda}$.

The temperature is
\begin{equation}
T=\frac{r_h}{\pi R^2}(1-\frac{Q^2}{2}),\label{T}
\end{equation}
with the restriction $0\leq Q\leq\sqrt{2}$ for the charge Q of the
black hole.

The chemical potential $\mu$ is given by
\begin{equation}
\mu=\frac{\sqrt{3}Qr_h}{R^2}.\label{mu}
\end{equation}

Note that in the limit $\mu\rightarrow0$, the usual
$AdS_5$-Schwarzschild metric is reproduced, as expected.

However, we should admit that the chemical potential implemented
in this way is not the quark (or baryon) chemical potential of QCD
but a chemical potential which is conjugate to an R-charge
associated with supersymmetry \cite{ED}. As it indeed behaves like
a chemical potential, one can use it as a simple way of
introducing the finite density effects into the system.

Next, to make the quark-antiquark pair moving, we assume that the
plasma is at rest and the frame is moving in one direction. Here
we boost the frame in the $x_3$ direction with rapidity $\beta$ so
that
\begin{equation}
dt=dt^\prime cosh\beta-dx_3^\prime sinh\beta, \qquad
dx_3=-dt^\prime sinh\beta+dx_3^\prime cosh\beta.\label{tr}
\end{equation}

Substituting (\ref{tr}) into (\ref{metric}) and dropping the
primes, we have the boosted metric as
\begin{eqnarray}
ds^2
=[-\frac{r^2}{R^2}f(r)cosh^2\beta+\frac{r^2}{R^2}sinh^2\beta]dt^2-2sinh\beta
cosh\beta[\frac{r^2}{R^2}-\frac{r^2}{R^2}f(r)]dtdx_3\nonumber\\+[-\frac{r^2}{R^2}f(r)sinh^2\beta+\frac{r^2}{R^2}cosh^2\beta]dx_3^2+\frac{r^2}{R^2}(dx_1^2+dx_2^2)+\frac{R^2}{r^2}f(r)^{-1}dr^2.\label{metric1}
\end{eqnarray}

\section{imaginary potential}

In this section, we follow the calculations of \cite{JN} to
analyze the imaginary potential with the metric (\ref{metric1}).
Generally, to investigate the moving quarkonium, one should
consider different alignments for the quarkonium with respect to
the plasma wind, including parallel $(\theta=0)$, transverse
$(\theta=\pi/2)$, or arbitrary direction $(\theta)$. In this
paper, we discuss the two extreme cases: $\theta=0$ and
$\theta=\pi/2$.

We now consider the system parallel to the wind in the $x_3$
direction, the coordinate is parameterized by
\begin{equation}
t=\tau, \qquad x_1=0,\qquad x_2=0,\qquad x_3=\sigma,\qquad
r=r(\sigma).\label{par}
\end{equation}

In this case, the quarks are located at $x_3=-\frac{L}{2}$ and
$x_3=\frac{L}{2}$, where $L$ is the inter-distance between the
$Q\bar{Q}$.

To proceed, the Nambu-Goto action is
\begin{equation}
S=-\frac{1}{2\pi\alpha^\prime}\int d\tau d\sigma\mathcal
L=-\frac{1}{2\pi\alpha^\prime}\int d\tau d\sigma\sqrt{-g},
\label{S}
\end{equation}
here $g$ is the determinant of the induced metric with
\begin{equation}
g_{\alpha\beta}=g_{\mu\nu}\frac{\partial
X^\mu}{\partial\sigma^\alpha} \frac{\partial
X^\nu}{\partial\sigma^\beta},
\end{equation}
where $g_{\mu\nu}$ is the metric, $X^\mu$ is the target space
coordinates. $\sigma^\alpha$ parameterize the world sheet with
$\alpha=0,1$.

From (\ref{metric1}) and (\ref{par}), we have the induced metric
\begin{equation} g_{00}=\frac{r^2}{R^2}f(r)cosh^2\beta-\frac{r^2}{R^2}sinh^2\beta, \qquad
g_{11}=\frac{r^2}{R^2}cosh^2\beta-\frac{r^2}{R^2}f(r)sinh^2\beta+\frac{R^2}{f(r)r^2}\dot{r}^2.
\end{equation}
with $\dot{r}=dr/d\sigma$.

Then the lagrangian density can be written as
\begin{equation}
\mathcal L=\sqrt{a(r)+b(r)\dot{r}^2}.
\end{equation}
where
\begin{eqnarray}
&a(r)&=\frac{r^4}{R^4}[f(r)sinh^4\beta+f(r)cosh^4\beta-sinh^2\beta
cosh^2\beta(1+f^2(r))],\nonumber\\&b(r)&=cosh^2\beta-\frac{1}{f(r)}sinh^2\beta.
\end{eqnarray}

As the action does not depend on $\sigma$ explicitly, the solution
satisfies
\begin{equation}
\mathcal L-\frac{\partial\mathcal
L}{\partial\dot{r}}\dot{r}=constant.
\end{equation}

The deepest point of the U-shaped string is $r=r_c$ with
$\dot{r}=0$, one gets
\begin{equation}
\dot{r}=\frac{dr}{d\sigma}=\sqrt{\frac{a^2(r)-a(r)a(r_c)}{a(r_c)b(r)}}\label{dotr}.
\end{equation}
with
\begin{eqnarray}
a(r_c)&=&\frac{r_c^4}{R^4}[f(r_c)sinh^4\beta+f(r_c)cosh^4\beta-sinh^2\beta
cosh^2\beta(1+f^2(r_c))],\nonumber\\
f(r_c)&=&1-(1+Q^2)(\frac{r_h}{r_c})^4+Q^2(\frac{r_h}{r_c})^6.
\end{eqnarray}

Integrating (\ref{dotr}), one finds the separate length of the
$Q\bar{Q}$ as
\begin{equation}
L=2\int_{r_c}^{\infty}dr\sqrt{\frac{a(r_c)b(r)}{a^2(r)-a(r)a(r_c)}}\label{x}.
\end{equation}

The real part of the heavy quark potential can be derived from the
classical solution of the equations of motion
\begin{equation}
ReV_{Q\bar{Q}}=\frac{1}{\pi\alpha^\prime}\int_{r_c}^{\infty}dr[\sqrt{\frac{a(r)b(r)}{a(r)-a(r_c)}}-\sqrt{b(r_0)}]-
\frac{1}{\pi\alpha^\prime}\int_{r_h}^{r_c}dr\sqrt{b(r_0)}\label{re},
\end{equation}
where $b(r_0)=b(r\rightarrow\infty)$.

The imaginary part of the heavy quark potential can be obtained by
using the thermal worldsheet fluctuation method
\begin{equation}
ImV_{Q\bar{Q}}=-\frac{1}{2\sqrt{2}\alpha^\prime}[\frac{a^\prime(r_c)}{2a^{\prime\prime}(r_c)}-\frac{a(r_c)}{a^\prime(r_c)}]\sqrt{b(r_c)},\label{im}
\end{equation}
where the derivations are with respect to $r$. Note that one can
recover the Im$V_{Q\bar{Q}}$ of the moving quarkonium \cite{MAL}
by taking $\mu=0$ in (\ref{im}).

Now, we are ready to calculate the inter-distance and imaginary
potential. From (\ref{T}), (\ref{mu}) and (\ref{x}), the resulting
expression for $LT$ can be written as
\begin{equation}
LT=2\frac{r_h}{\pi R^2}(1-\frac{({\frac{\mu
R^2}{\sqrt{3}r_h})}^2}{2})\int_{r_c}^{\infty}dr\sqrt{\frac{a(r_c)b(r)}{a^2(r)-a(r)a(r_c)}}\label{LT}.
\end{equation}

\begin{figure}
\centering
\includegraphics[width=8cm]{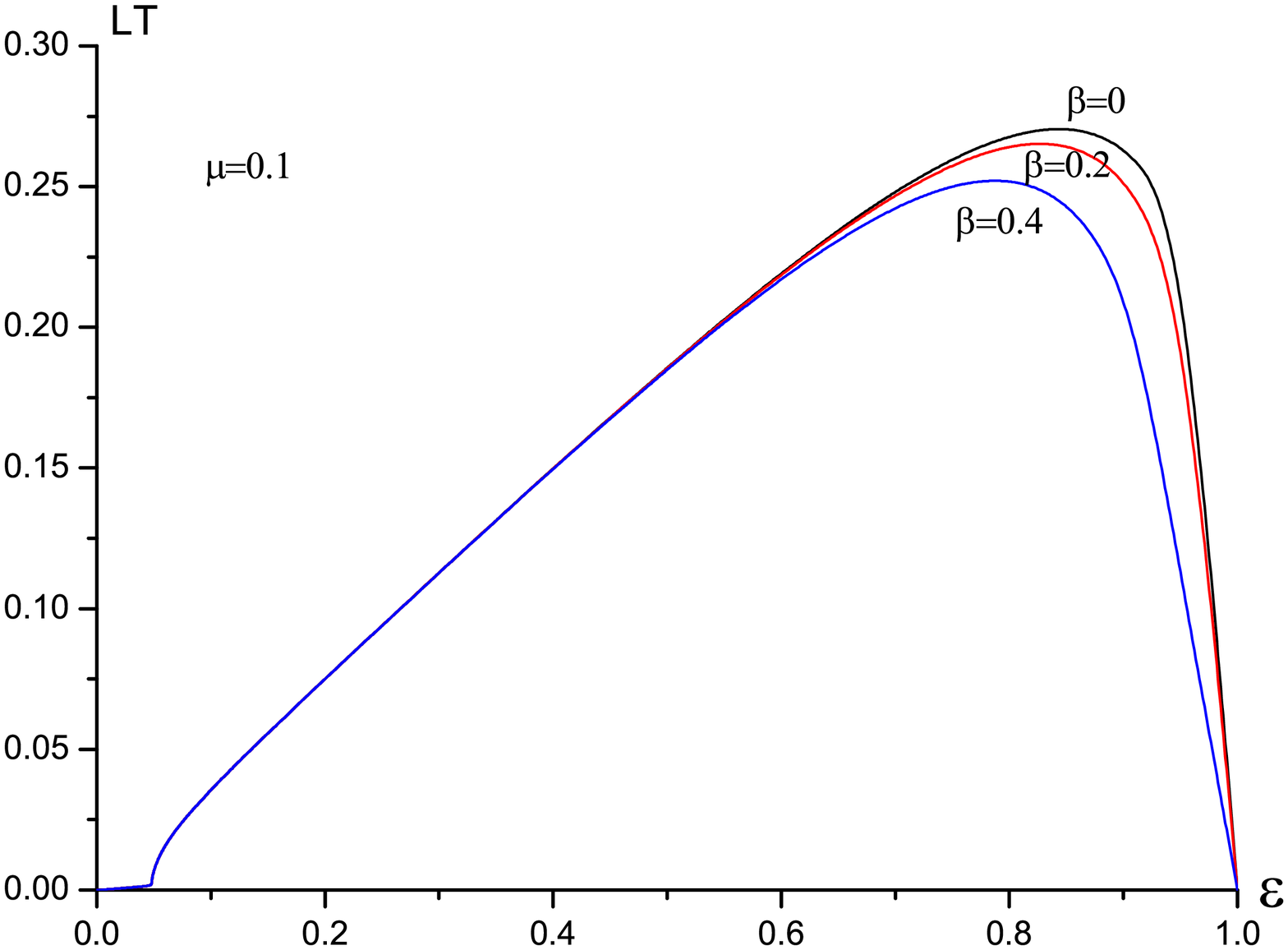}
\includegraphics[width=8cm]{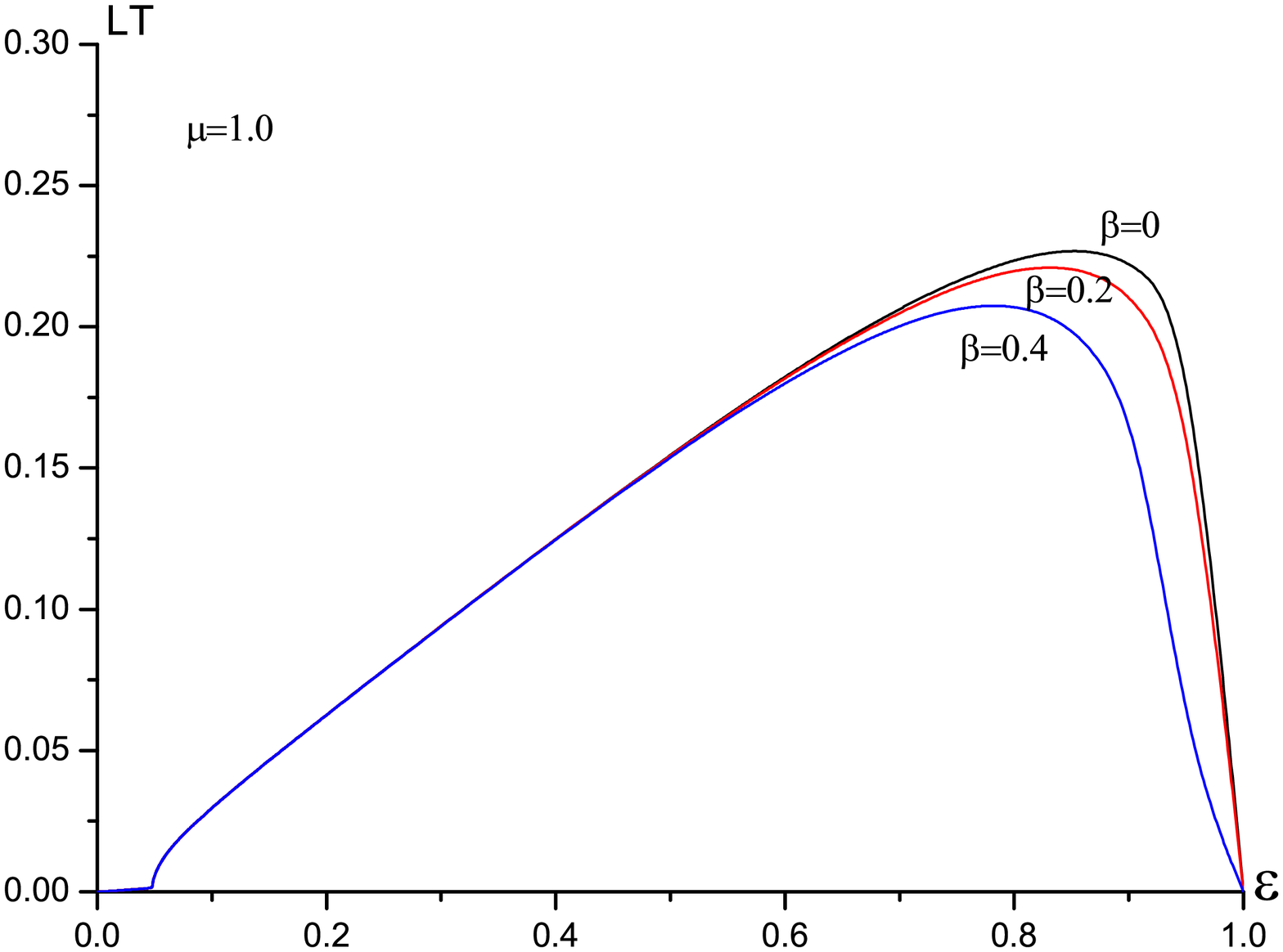}
\caption{$LT$ versus $\varepsilon$ for $\theta=0$. Left:
$\mu=0.1$. Right: $\mu=1$. In all of the plots from top to bottom
$\beta=0,0.2,0.4$ respectively.}
\end{figure}

In fig.1, we plot $LT$ as a function of $\varepsilon$ with
$\varepsilon\equiv r_h/r_c$ for various cases. One can see that
for each plot there is a maximum value of $\varepsilon_{max}$, and
that $LT$ is a increasing function of $\varepsilon$ for
$\varepsilon<\varepsilon_{max}$. On the other hand, for
$LT>LT_{max}$, $LT$ is a decreasing function of $\varepsilon$.
Actually, in this case one needs to take into account new
configurations \cite{DB} which are not solutions of the Nambu-Goto
action. Here we consider only $LT<LT_{max}$.

In fig.1 we show $LT$ versus $\varepsilon$ for $\theta=0$ of three
different rapidity. The left panel of fig.1 is plotted for a small
value of chemical potential ($\mu=0.1$) while the right panel is
for a larger value of chemical potential ($\mu=1.0$). We can see
that at a fixed $\mu$ increasing the rapidity leads to decreasing
the $LT_{max}$. This result is in agreement with that in
\cite{MAL,KB1}. In addition, by comparing the left panel with the
right panel, we can see that for each $\beta$ as $\mu$ increases
the $LT_{max}$ decreases. In other words, the presence of the
chemical potential tends to decrease the $LT_{max}$.

Then we calculate the imaginary potential from (\ref{im}), after
some algebra one finds
\begin{eqnarray}
a^\prime(r)=(sinh^4\beta+cosh^4\beta)(4r^3f(r)+r^4f^\prime(r))-sinh^2\beta
cosh^2\beta(4r^3+4r^3f^3(r)+2r^4f(r)f^\prime(r))\label{1},
\end{eqnarray}
\begin{eqnarray}
a^{\prime\prime}(r)=(sinh^4\beta+cosh^4\beta)(12r^2f(r)+8r^3f^\prime(r)+r^4f^{\prime\prime}(r))-sinh^2\beta
cosh^2\beta\nonumber\\
(12r^2+12r^2f^2(r)+8r^3f(r)f^\prime(r)+2r^4f(r)f^{\prime\prime}(r)+8r^3f^\prime(r)f(r)+2r^4f^\prime(r)f^\prime(r))\label{2},
\end{eqnarray}
with
\begin{eqnarray}
f^\prime(r)=4(1+Q^2)r_h^4r^{-5}-6Q^2r_h^6r^{-7},\qquad
f^{\prime\prime}(r)=-20(1+Q^2)r_h^4r^{-6}+42Q^2r_h^6r^{-8}.\label{3}
\end{eqnarray}

From (\ref{im}), (\ref{1}), (\ref{2}) and (\ref{3}), we can obtain
the $ImV_{Q\bar{Q}}$. Numerically, we plot $ImV/(\sqrt{\lambda}T)$
against $LT$ for $\theta=0$ of three different rapidity in fig.2.
For each plot we can see that the imaginary potential starts at a
$L_{min}$ which can be found by solving $ImV_{Q\bar{Q}}=0$ and
ends at a $L_{max}$. In addition, at a fixed chemical potential,
by increasing the rapidity the absolute value of the imaginary
potential decreases. Moreover, comparing the left panel with the
right one, one finds increasing the chemical potential leads to
decreasing the absolute value of the imaginary potential. In other
words, turning on the chemical potential effect leads to
generating the $ImV_{Q\bar{Q}}$ for smaller inter-quark distances.
It was argued \cite{JN} that the imaginary potential can be used
to estimate the thermal width of heavy quarkonium and in general a
large thermal width corresponding to a large dissociation length.
Thus, the chemical potential has the effect of decreasing the
thermal width or decreasing the dissociation length.
Interestingly, the higher derivative corrections has the similar
behavior \cite{KB1}.
\begin{figure}
\centering
\includegraphics[width=8cm]{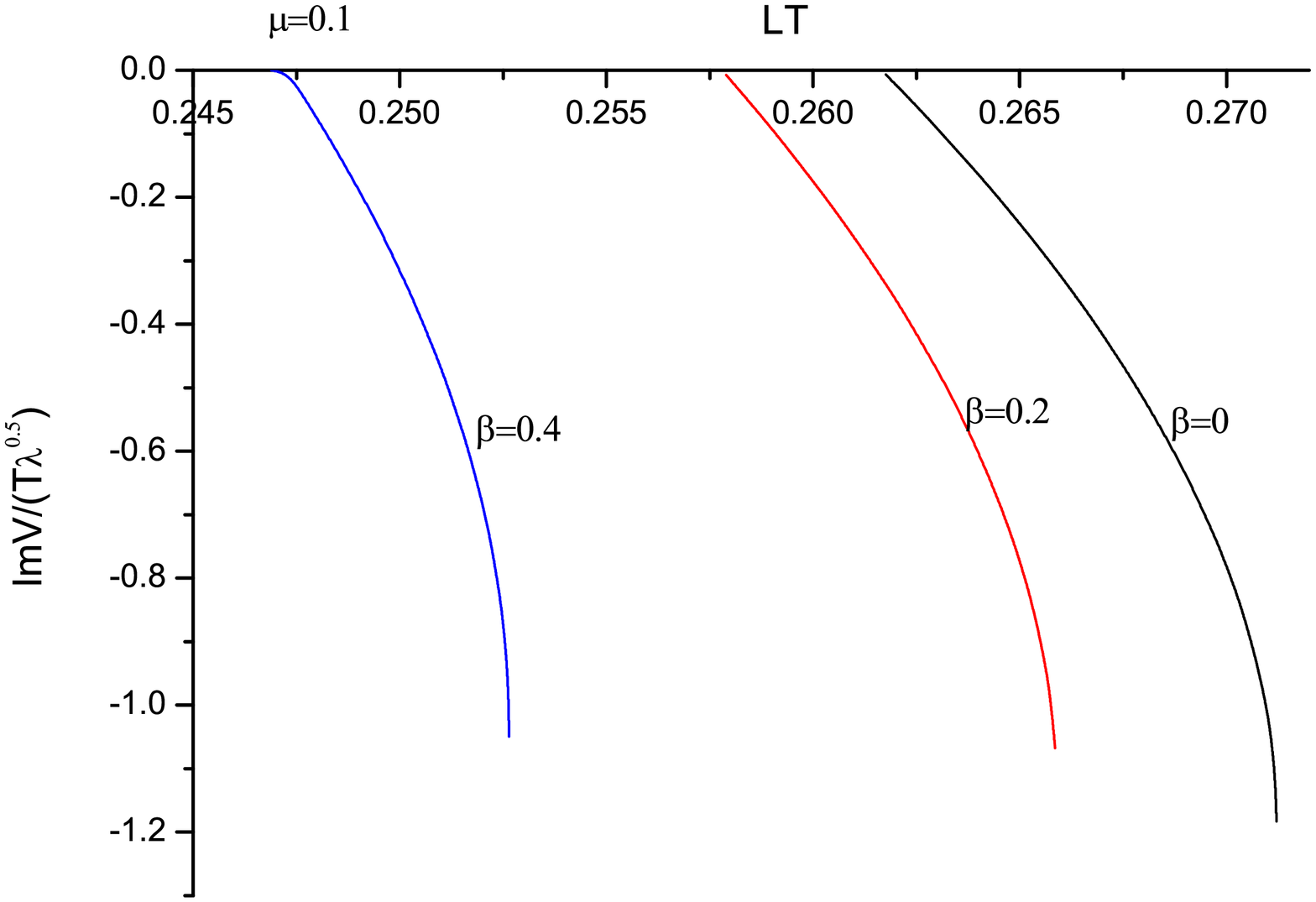}
\includegraphics[width=8cm]{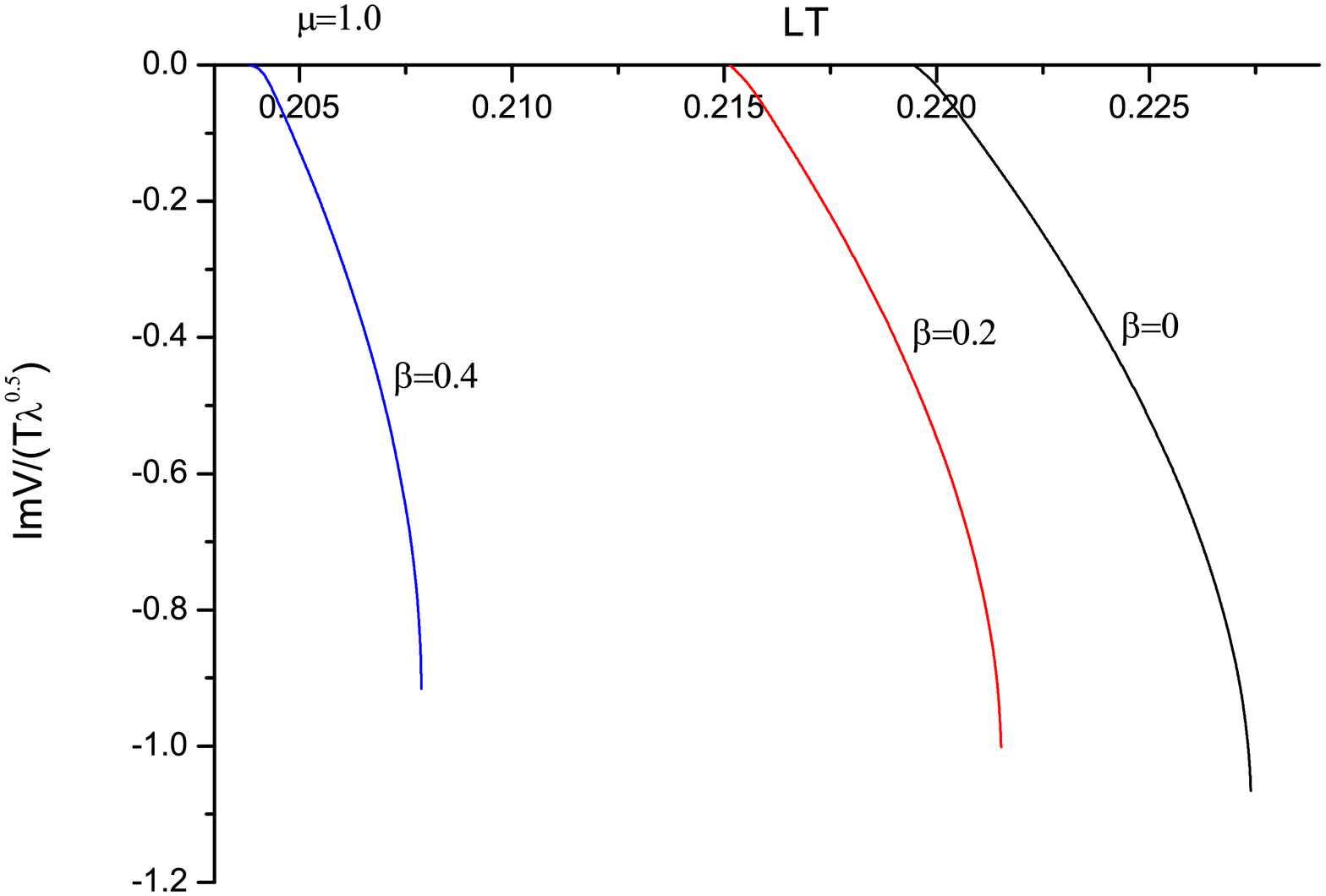}
\caption{$ImV/(\sqrt{\lambda}T)$ versus $LT$ for $\theta=0$. Left:
$\mu=0.1$. Right: $\mu=1$. In all of the plots from left to right
$\beta=0.4,0.2,0$ respectively.}
\end{figure}

Next, we consider the system transverse to the wind in the $x_1$
direction. The parametrization is
\begin{equation}
t=\tau, \qquad x_1=\sigma,\qquad x_2=0,\qquad x_3=0,\qquad
r=r(\sigma).\label{par1}
\end{equation}
here the quarks are located at $x_1=-\frac{L}{2}$ and
$x_1=\frac{L}{2}$.

The next analysis is similar to the parallel case in the previous
section. So we here focus on the results.

\begin{figure}
\centering
\includegraphics[width=8cm]{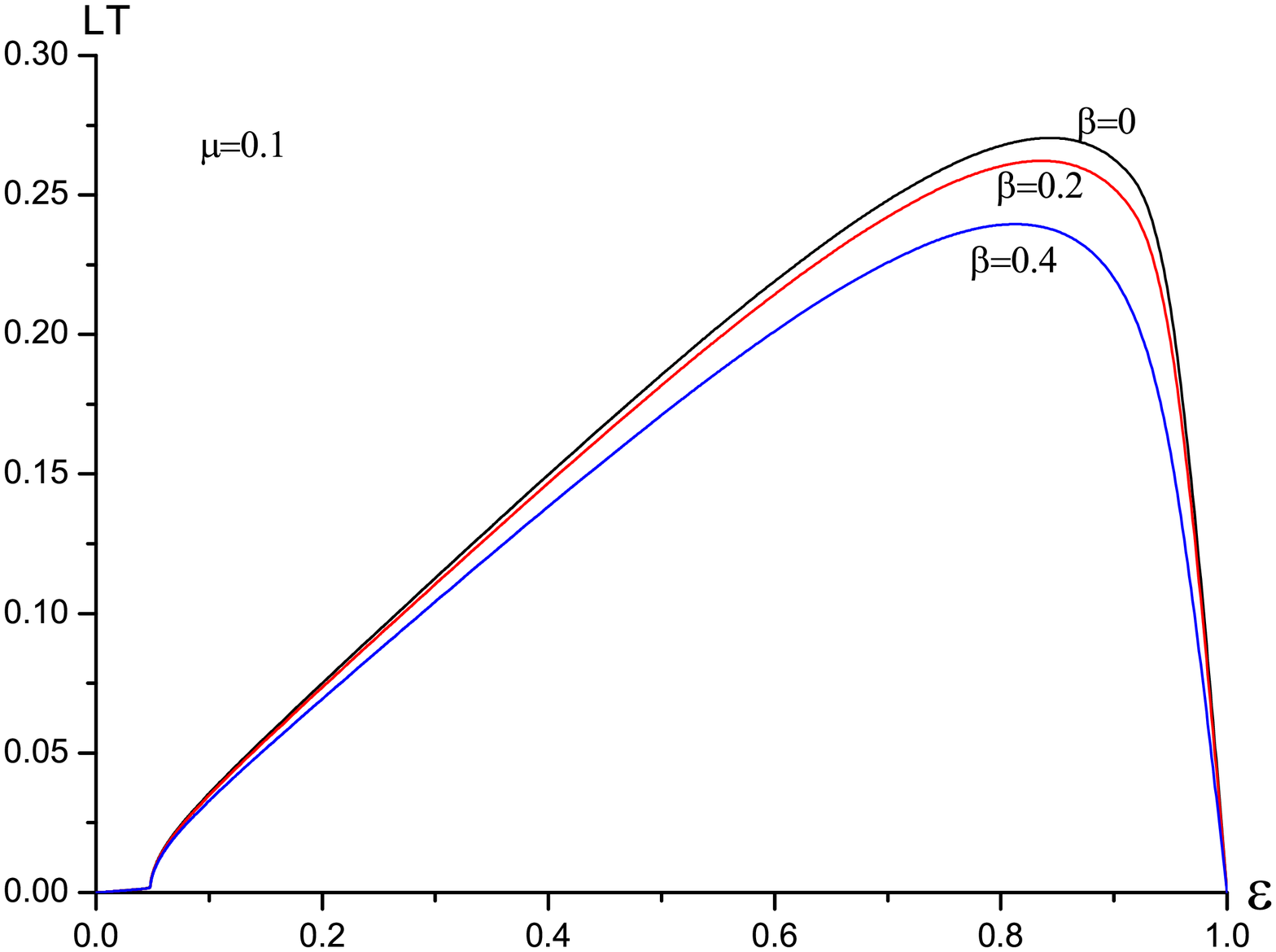}
\includegraphics[width=8cm]{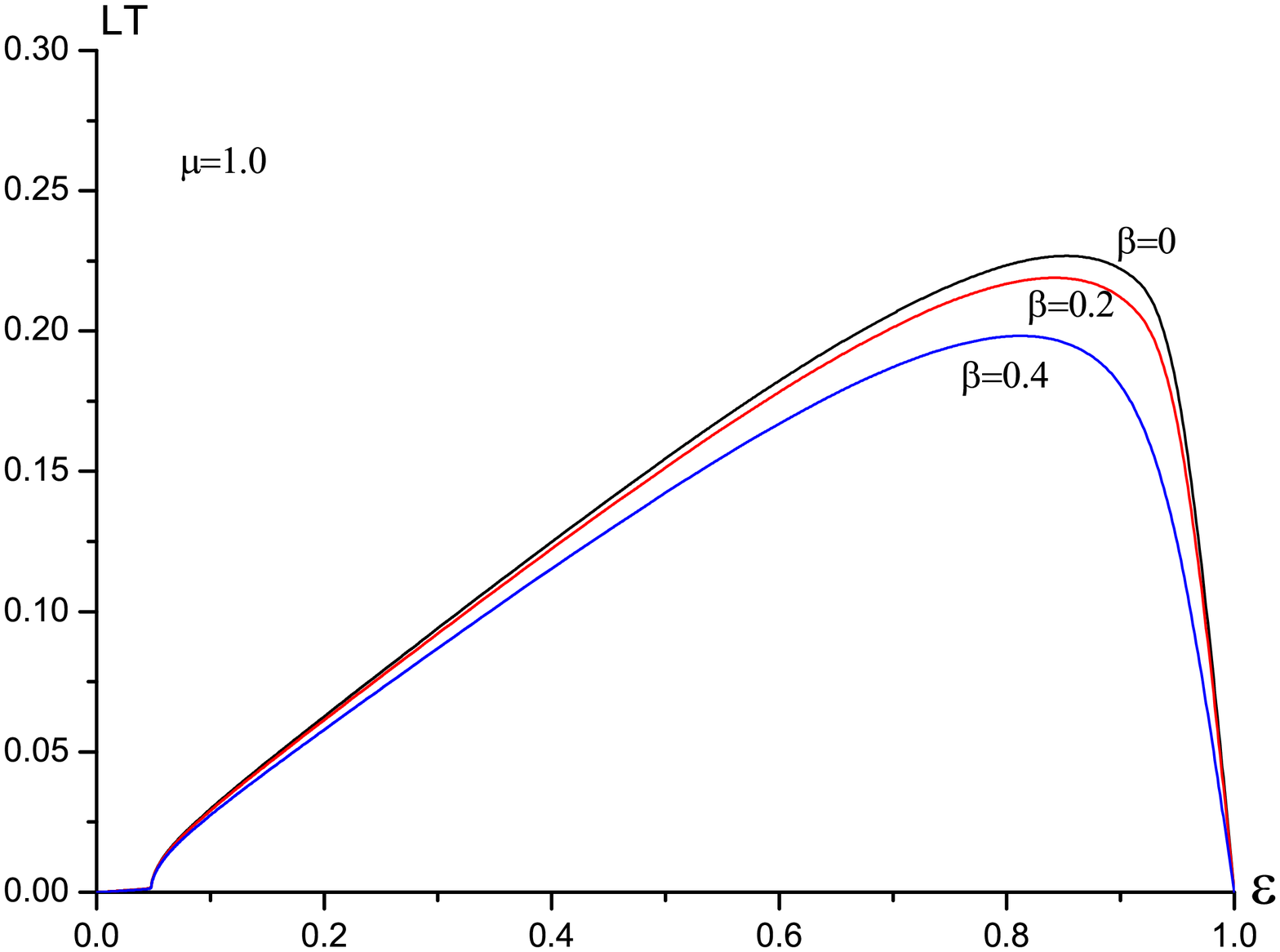}
\caption{$LT$ versus $\varepsilon$ for $\theta=\pi/2$. Left:
$\mu=0.1$. Right: $\mu=1$. In all of the plots from top to bottom
$\beta=0,0.2,0.4$ respectively.}
\end{figure}

\begin{figure}
\centering
\includegraphics[width=8cm]{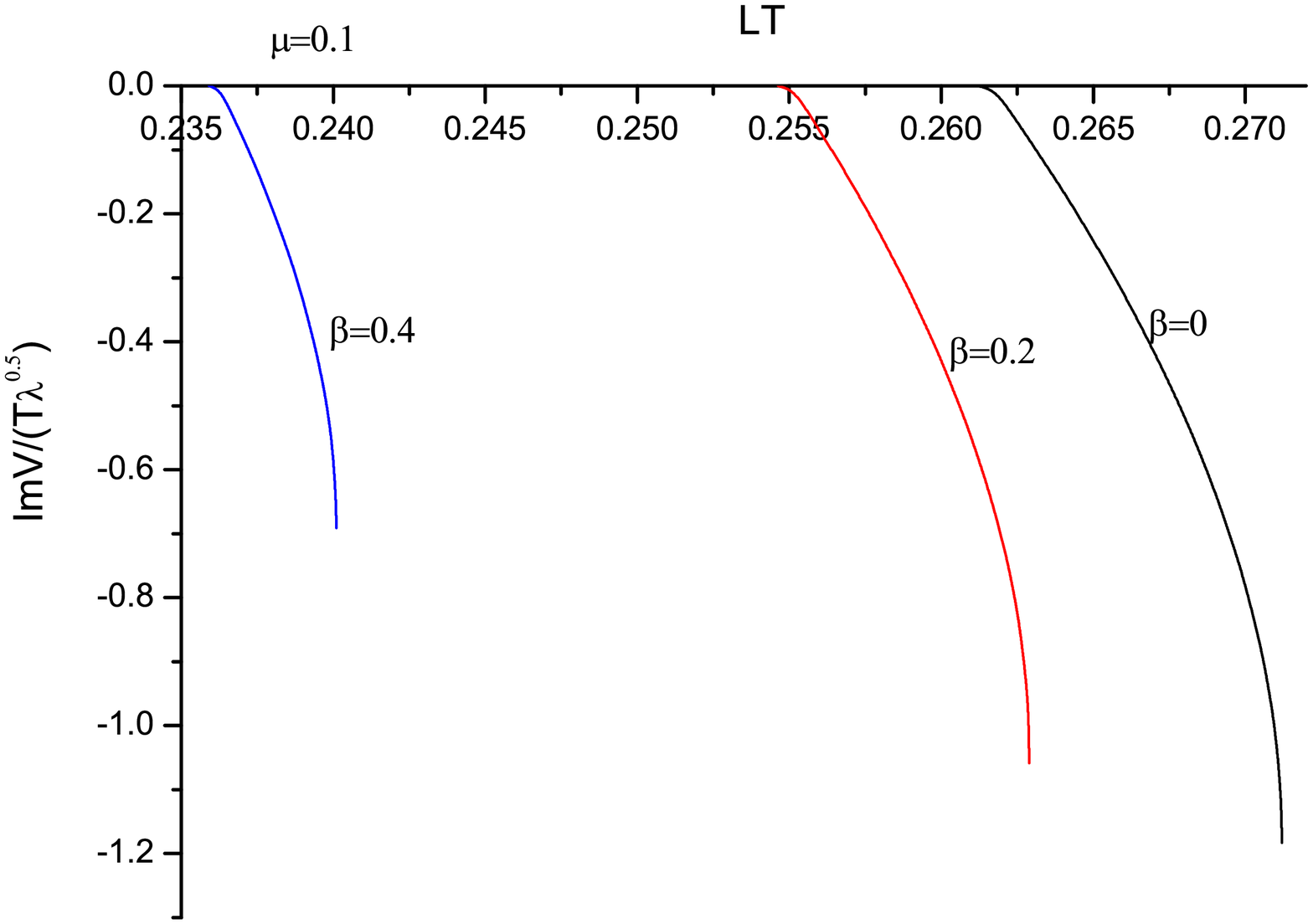}
\includegraphics[width=8cm]{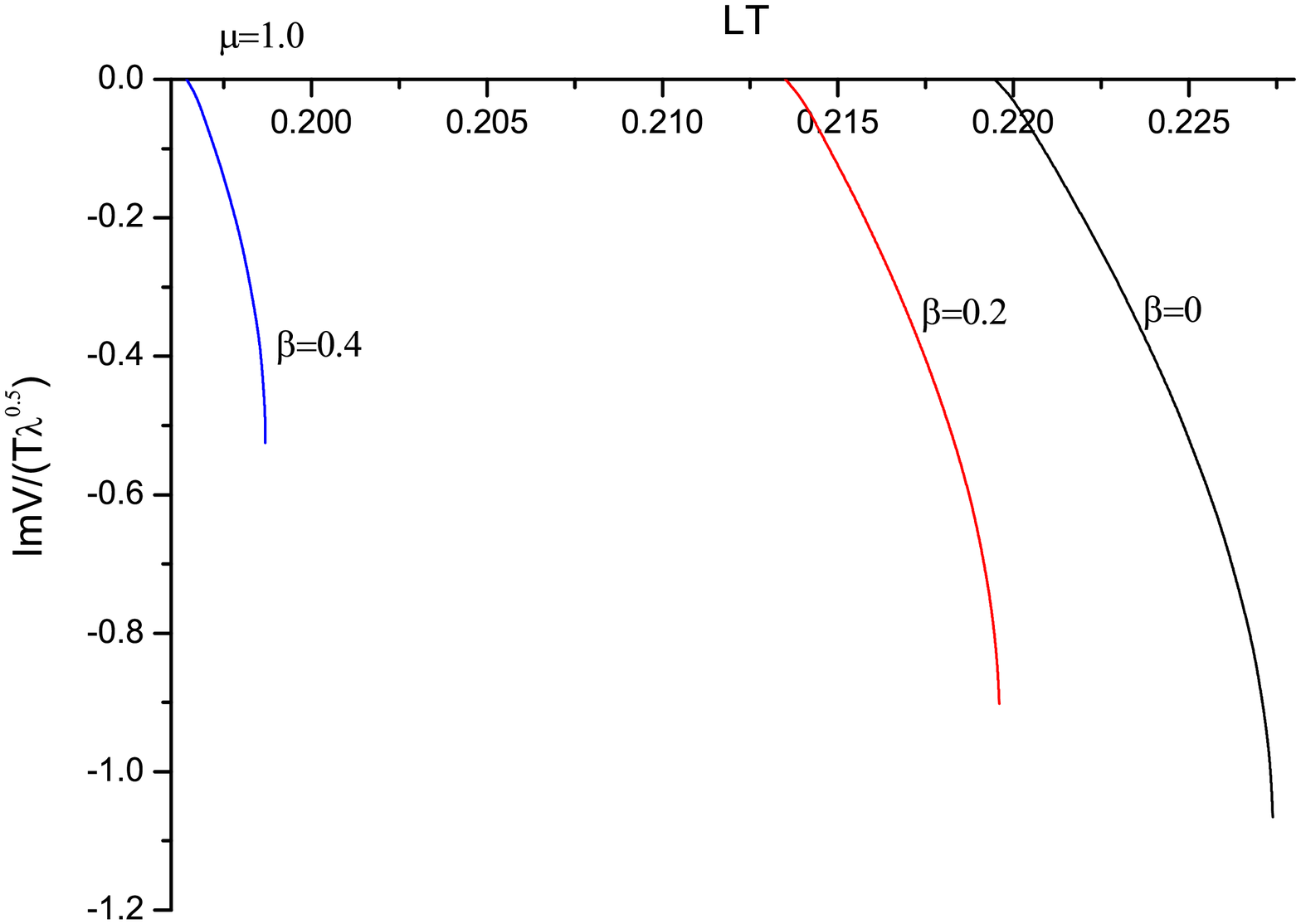}
\caption{$ImV/(\sqrt{\lambda}T)$ versus $LT$ for $\theta=\pi/2$.
Left: $\mu=0.1$. Right: $\mu=1$. In all of the plots from left to
right $\beta=0.4,0.2,0$ respectively.}
\end{figure}

In fig 3, the $LT$ versus $\varepsilon$ for $\theta=\pi/2$ at two
different chemical potential have been shown. In fig.4, the
$ImV/(\sqrt{\lambda}T)$ against $LT$ for $\theta=\pi/2$ at two
different chemical potential are also presented. One finds that
the results are very similar to the parallel case: At fixed
chemical potential, increasing the rapidity leads to decreasing
the absolute value of the imaginary potential; At fixed rapidity,
increasing the chemical potential also leads to decreasing the
absolute value of the imaginary potential. But one difference is
that stronger modification occurs at $\theta=\pi/2$ than that at
$\theta=0$. In other words, the chemical potential has important
effect when the quarkonium is moving transverse to the wind. This
behavior is also similar to that of the higher derivative
corrections \cite{KB1}.

We now conclude this section as follows:

1. The imaginary potential of the moving quarkonium depends on the
chemical potential.

2. At fixed chemical potential, as the rapidity increases the
absolute value of the imaginary potential decreases.

3. At fixed rapidity, as the chemical potential increases the
absolute value of the imaginary potential decreases.

4. The chemical potential has a larger effect when the quarkonium
is moving transverse to the wind than that parallel to the wind.

5. The presence of the chemical potential tends to decrease the
thermal width or the dissociation length.

\section{entropic force}
In this section, we analyze the chemical effect on the entropic
force with respect to a moving quarkonium with the metric
(\ref{metric1}).

In \cite{DEK}, the entropic force is proposed to be related to the
entropy S, that is
\begin{equation}
\mathcal{F}=T\frac{\partial S}{\partial L},\label{f}
\end{equation}
where $T$ is the temperature of the plasma, $L$ refers to the
inter-quark distance. Thus, to evaluate the entropic force, one
needs to calculate these quantities: $S,L$ and $T$, which have
been studied from the AdS/CFT correspondence \cite{JMM,ABR,SJR}.
In the next two section, we follow the calculation of \cite{KHA}.

For the case of $\theta=0$, the assumption and analysis are almost
parallel to the previous section. From the Eq.(\ref{T}) and
Eq.(\ref{x}), one finds
\begin{equation}
T=\frac{r_h}{\pi R^2}(1-\frac{Q^2}{2}),\qquad
L=2\int_{r_c}^{\infty}dr\sqrt{\frac{a(r_c)b(r)}{a^2(r)-a(r)a(r_c)}},
\end{equation}
with
\begin{eqnarray}
a(r_c)&=&\frac{r_c^4}{R^4}[f(r_c)sinh^4\beta+f(r_c)cosh^4\beta-sinh^2\beta
cosh^2\beta(1+f^2(r_c))],\nonumber\\
f(r_c)&=&1-(1+Q^2)(\frac{r_h}{r_c})^4+Q^2(\frac{r_h}{r_c})^6.
\end{eqnarray}
and
\begin{eqnarray}
&a(r)&=\frac{r^4}{R^4}[f(r)sinh^4\beta+f(r)cosh^4\beta-sinh^2\beta
cosh^2\beta(1+f^2(r))],\nonumber\\&b(r)&=cosh^2\beta-\frac{1}{f(r)}sinh^2\beta.
\end{eqnarray}

To study the entropy $S$, one can use the relation:
\begin{equation}
S=-\frac{\partial F}{\partial T},\label{s}
\end{equation}
where $F$ is the free energy of the $Q\bar{Q}$. Before studying
the free energy, we should pause here to stress some issues. From
the fig.1, we can see that for each plot $LT$ has a maximum value.
We call this value $c$. If $L>\frac{c}{T}$ the quarks are
completely screened while $L<\frac{c}{T}$ the fundamental string
is connected. Therefore, there are two cases for the free energy:

1. If $L>\frac{c}{T}$, the fundamental string will break in two
pieces implying that the quarks are screened. In this case, the
choice of the free energy $F^{(1)}$ is not unique \cite{MCH}, we
here opt for a configuration of two disconnected trailing drag
strings \cite{CPH}, that is
\begin{equation}
F^{(1)}=\frac{1}{\pi\alpha^\prime}\int_{r_h}^{\infty}dr.
\end{equation}

Applying (\ref{s}), one finds
\begin{equation}
S^{(1)}=\sqrt{\lambda}\theta(L-\frac{c}{T})\label{S2}.
\end{equation}

Note that $c$ is a decreasing function of $\mu$ and $\beta$. One
can obtain the value of $c$ with numerical methods and also one
can get it from the fig.1.

2. If $L<\frac{c}{T}$, the free energy of the $Q\bar{Q}$ can be
derived from the on-shell action of the fundamental string in the
dual geometry. Substituting (\ref{dotr}) into (\ref{S}), one gets
\begin{equation}
F^{(2)}=\frac{1}{\pi\alpha^\prime}\int_{r_c}^{\infty} dr
\sqrt{\frac{a(r)b(r)}{a(r)-a(r_c)}}.
\end{equation}

Then using the (\ref{s}) one can calculate the entropy $S^{(2)}$
numerically. In fig.5, we plot $S^{(2)}/\sqrt{\lambda}$ versus
$LT$ at two different $\mu$ for various $\beta$. The left panel is
plotted for $\mu=0.1$ while the right one is plotted for
$\mu=1.0$. From the plots, we can see that at a fixed chemical
potential by increasing the rapidity the entropy increases. This
is consistent with that in \cite{KBF}. In addition, to see the
chemical effect, we compare the left panel with the right one and
find that increasing $\mu$ leads to increasing the entropy. As
stated above, the entropic force, responsible for dissociating the
quarkonium, is related to the growth of the entropy with the
distance. Therefore, increasing $\mu$ leads to increasing the
entropic force or decreasing the dissociation length.
Interestingly, the medium effect on the entropic force has also
been studied in some other backgrounds \cite{KBF}. The behavior
there is similar.

\begin{figure}
\centering
\includegraphics[width=8cm]{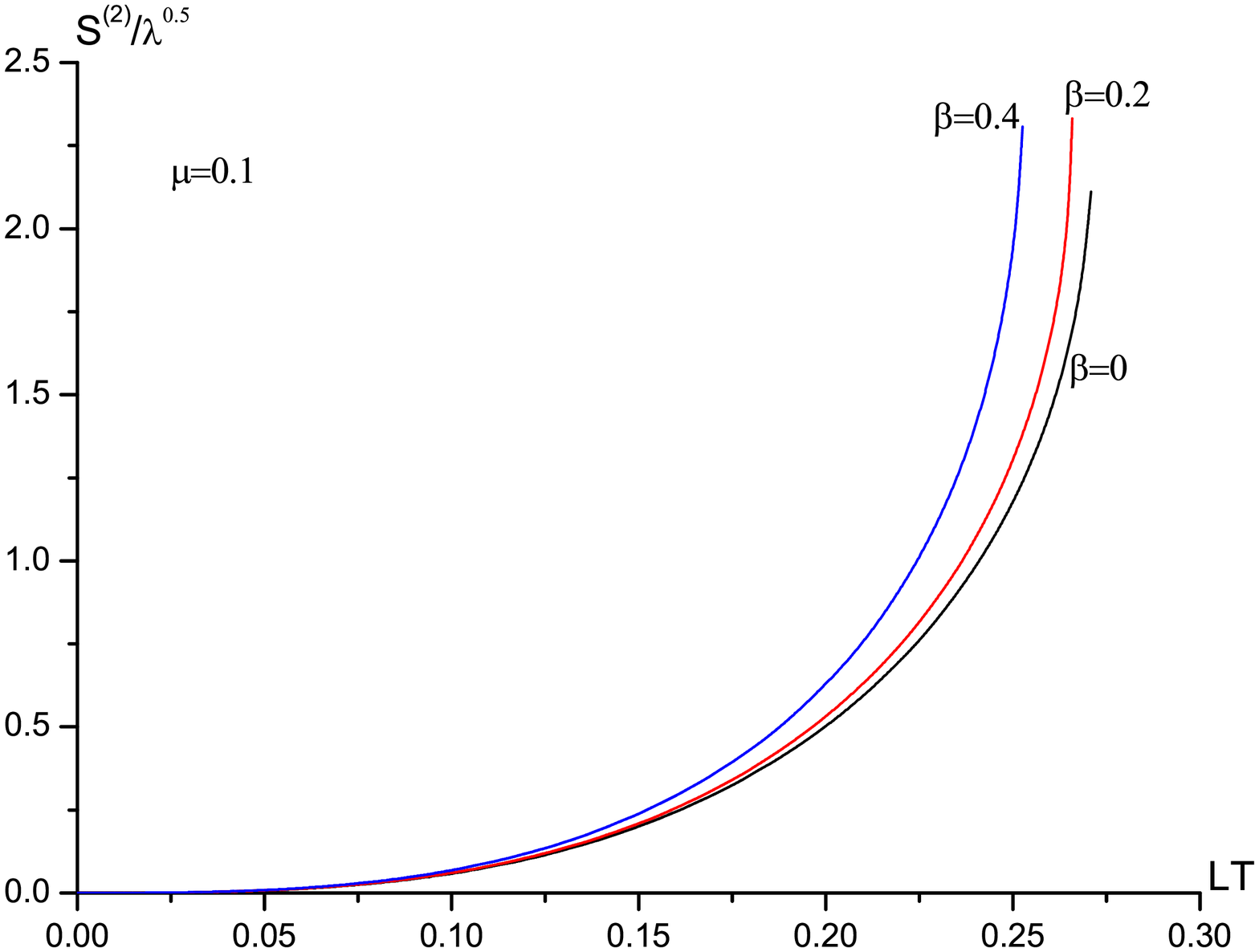}
\includegraphics[width=8cm]{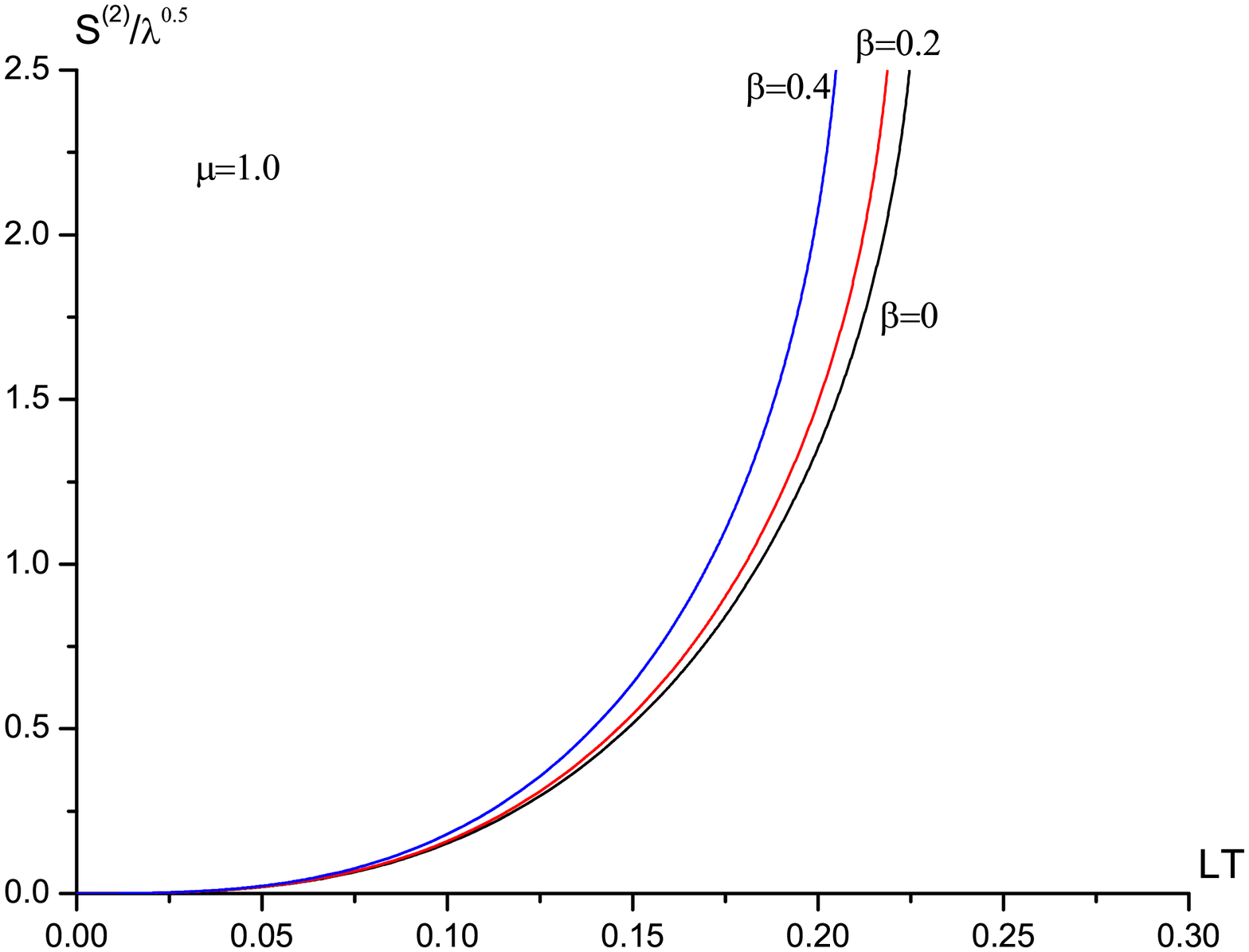}
\caption{$S^{(2)}/\sqrt{\lambda}$ versus $LT$ for $\theta=0$.
Left: $\mu=0.1$. Right: $\mu=1$. In all of the plots from left to
right $\beta=0.4,0.2,0$ respectively.}
\end{figure}

For the case of $\theta=\pi/2$, we concern mainly with the final
plots. In fig.6, we plot $S^{(2)}/\sqrt{\lambda}$ as a function of
$LT$ for $\theta=\pi/2$ at two chemical potential. We observe that
the behavior is almost the same as the case of $\theta=0$. The
only difference is that $\mu$ has large effect to the entropic
force when the quarkonium is moving transverse to the wind
comparing with the parallel motion.

\begin{figure}
\centering
\includegraphics[width=0.4\textwidth,bb=0 0 800 600]{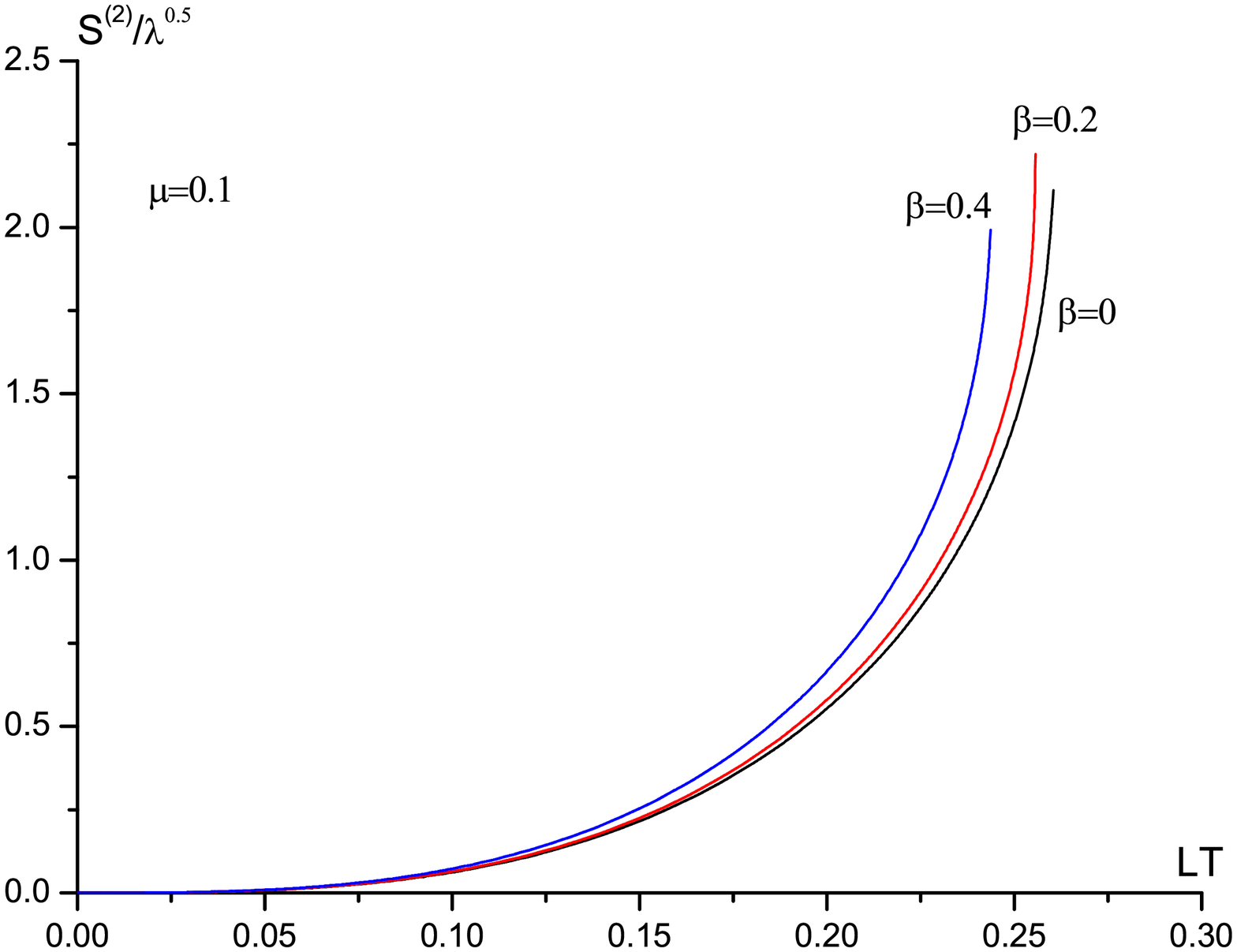}
\includegraphics[width=0.4\textwidth,bb=0 0 800 600]{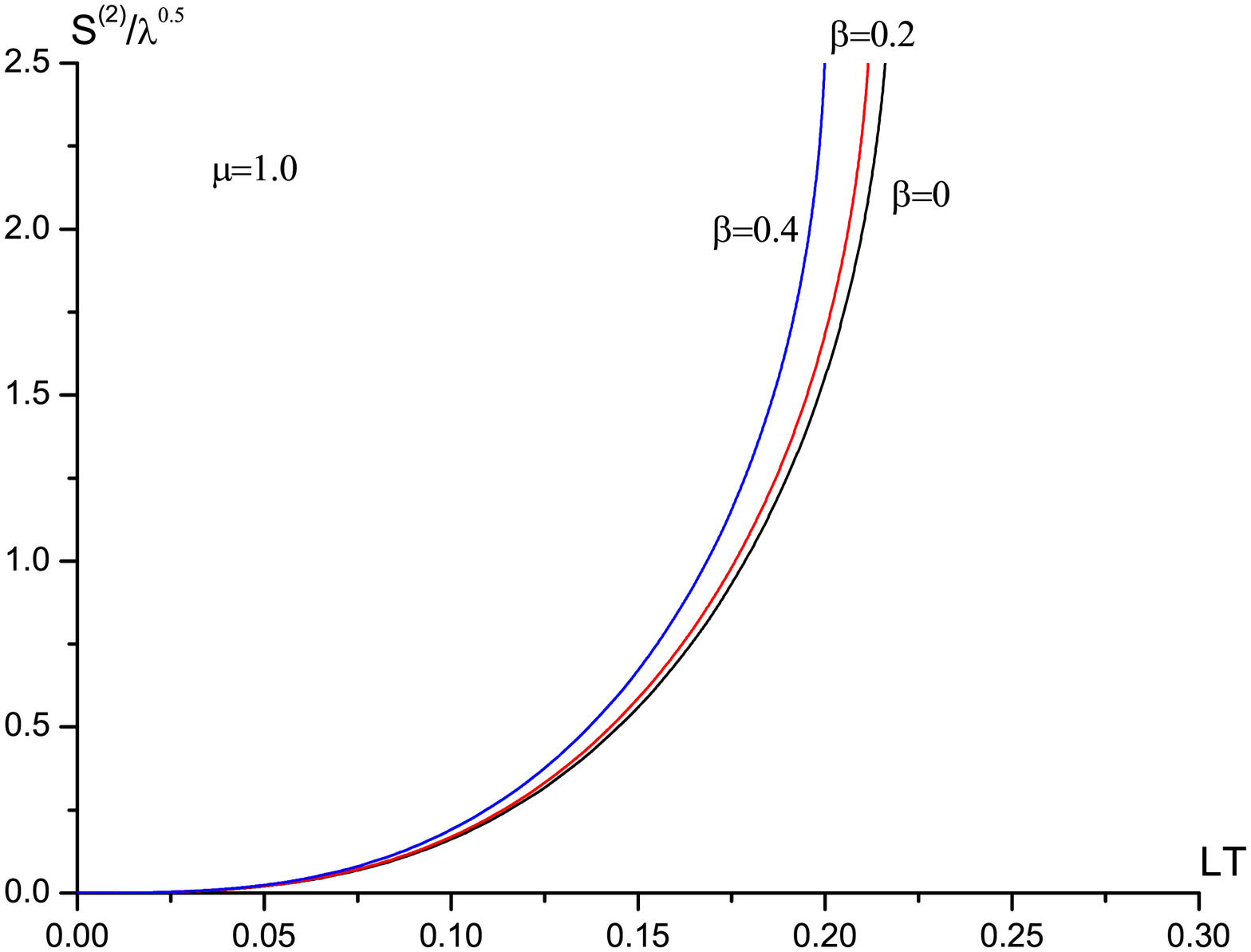}
\caption{$S^{(2)}/\sqrt{\lambda}$ versus $LT$ for $\theta=\pi/2$.
Left: $\mu=0.1$. Right: $\mu=1$. In all of the plots from left to
right $\beta=0.4,0.2,0$ respectively.}
\end{figure}

We now conclude this section as follows:

1. The entropic force with respect to a moving quarkonium depends
on the chemical potential.

2. At fixed chemical potential, as the rapidity increases the
entropic force increases.

3. At fixed rapidity, as the chemical potential increases the
entropic force increases.

4. The chemical potential has a larger effect to the entrpic force
when the quarkonium is moving transverse to the wind than that
parallel to the wind.

5. The presence of the chemical potential tends to increase the
entropic force or decrease the dissociation length.

\section{conclusion}
In heavy ion collisions at LHC and RHIC, the quarkonium is usually
moving through the sQQP with relativistic velocities. An
understanding of how the quarkonium affected by the medium may be
essential for theoretical predictions. Recently, the imaginary
potential and the entropic force have been proposed to responsible
for melting the heavy quarkonium respectively. The two quantities
represent different mechanisms, so it is of interest to compare
them. Evaluations of a same effect to these quantities could be
considered as a simple comparison.

In this paper, we have investigated the chemical potential effect
on the imaginary potential and the entropic force associated with
a moving quarkina from the AdS/CTF duality. It is shown that for
both mechanisms the chemical potential has the same effect: For
the imaginary potential mechanism, the presence of the chemical
potential tends to decrease the thermal width or decrease the
dissociation length. For the entropic force mechanism, the
chemical potential has the effect of increasing the entropic force
or decreasing the dissociation length. In other words, for both
mechanisms, it is found that the moving quarkonium dissociates
easier at finite density.

However, the deep connection between the two mechanisms is still
unknown. We hope to report our progress in this regard in future.

\section{Acknowledgments}

This research is partly supported by the Ministry of Science and
Technology of China (MSTC) under the ¡°973¡± Project no.
2015CB856904(4). Zi-qiang Zhang is supported by the NSFC under
Grant no. 11547204. Gang Chen is supported by the NSFC under Grant
no. 11475149. De-fu Hou is partly supported by the NSFC under Grant
nos. 11375070, 11221504 and 11135011.


\end{document}